# Isolated anions induced high ionic conductivity

Qifan Yang[1,2], Jing Xu[1,3], Yuqi Wang[1,3], Xiao Fu[1,2], Ruijuan Xiao[1,2,3]* and Hong Li[1,2,3]*

[1] Beijing National Laboratory for Condensed Matter Physics, Institute of Physics, Chinese Academy of Sciences, Beijing 100190, China.

[2] Center of Materials Science and Optoelectronics Engineering, University of Chinese Academy of Sciences, Beijing 100049, China

[3] School of Physical Sciences, University of Chinese Academy of Sciences, Beijing 100049, China

E-mail: rjxiao@iphy.ac.cn, hli@iphy.ac.cn

# Abstract

One of the key materials in solid-state lithium batteries is fast ion conductors. However, the $Li^+$ ion transport in inorganic crystals involves complex factors, making it a mystery to find and design ion conductors with low migration barriers. In this work, a distinctive structural characteristic involving isolated anions has been discovered to enhance high ionic conductivity in crystals. It is an effective way to create a smooth energy potential landscape and construct local pathways for lithium ion migration. By adjusting the spacing and arrangement of the isolated anions, these local pathways can connect with each other, leading to high ion conductivity. By designing different space groups and local environments of the $Se^{2-}$ anions in the $Li_8SiSe_6$ composition, combined with the ion transport properties obtained from AIMD simulations, we define isolated anions and find that local environment with higher point group symmetry promotes the formation of cage-like local transport channels. Additionally, the appropriate distance between neighboring isolated anions can create coplanar connections between adjacent cage-like channels. Furthermore, different types of isolated anions can be used to control the distribution of cage-like channels in the lattice. Based on the structural characteristic of isolated anions, we discovered compounds with isolated $N^{3-}$, $Cl^-$, $I^-$, and $S^{2-}$ features from the crystal structure databases. The confirmation of ion transport in these structures validates the proposed design method of using isolated anions as structural features for fast ion conductors and leads to the discovery of several new fast ion conductor materials.

# Introduction

Nowadays, lithium ion batteries are becoming increasingly popular throughout the fast growing markets of mobile electronic devices, electric vehicles and energy storage technologies.[1,2] Compared with currently used lithium ion batteries based on organic liquid electrolytes, all-solid-state lithium batteries have caught more and more attention.[3-6] In all-solid-state lithium batteries, the design of solid-state electrolytes is crucial[7]. Up to now, some solid electrolytes have been discovered and well explored. Li argyodites[8,9], $Li_{10}GeP_2S_{12}$ (LGPS)[10,11], NASICON-type $LiM_2(PO_4)_3$ (M = Ge, Ti, Sn, Hf, Zr)[12], garnet-type $Li_xLa_3M_2O_{12}$ ($5 \leq x \leq 7$, M = Nb, Ta, Sb, Zr, Sn)[13,14], halide-based Li-M-X (M = metal element, X = F, Cl, Br, I)[15,16] and so on exhibit ionic conductivity similar or close to that of liquid electrolytes, but the mystery of the relationship between lattice atomic structure and ionic conductivity remains unresolved. Lots of studies attempt to summarize the characteristics of current fast ion conductors[17,18] and try to discover more high-performance solid-state electrolytes[19,20]. Despite the property of mobile cations in the lattice, the function of the framework consisting of immobile ions offers another perspective for research. Jun et al.[21] claimed that corner-sharing connectivity of the oxide crystal structure framework promotes superionic conductivity. Wang et al.[18] revealed the structural feature of face-sharing high-coordination sites for fast sodium-ion conductors. Furthermore, similar studies of the characteristic of anions in framework structures have been made. Wang et al.[17] believed that a structure with a bcc anion lattice is more likely to show high ionic conductivity, and the closer the anion lattice is to a perfect bcc lattice, the higher the conductivity. Treating all anions as an entirety is beneficial for identifying their spatial connectivity features, but overlooks their specific features in the local environment. In this work, we identified a special type of anions which do not bond with immobile cations in the lattice, but only form weak bonds with moving lithium ions. We refer to them as isolated anions. We found that the presence of isolated anions can lead to the frustration phenomenon and facilitate Li$^+$ ion transport by forming a flat potential energy surface (PES) around themselves to reduce the Li$^+$ migration barriers. The typical example is lithium argyrodite Li$_6$PS$_5$X (X = Cl, Br)[22], in which the anion lattice is close to the fcc configuration and deviate far from the bcc one but show ionic conductivity as high as 1.1×10$^{-3}$ S/cm[23]. Further looking into the lattice, two different types of anions with distinct chemical environments — isolated anions and bonded anions, can be found. As shown in Figure 1a, isolated anions (S$_{iso}$, Cl$_{iso}$) are considered as the S or Cl anions those do not form chemical bonding with P atoms but only bond with Li ions, and they occupy Wyckoff positions 4d and 4a. On the other hand, bonded anions (S$_{bond}$) refer to the S anions that form PS$_4$ tetrahedra by bonding with P atoms, and they occupy Wyckoff position 16e. The different local structure of S$_{iso}$, S$_{bond}$ and Cl$_{iso}$, as well as their various partial electronic density of states (PDOS) in Figure 1c demonstrate the inconsistency of their chemical environments. In previous studies, the transport of Li$^+$ in argyrodites was divided into three types of events, named as doublet jump, intra-cage hoppings and inter-cage hoppings[24]. According to the Li$^+$ diffusion pathways in Li$_6$PS$_5$Cl revealed by AIMD simulations shown as Figure 1b, the isolated anions S$_{iso}$ play a role as the center of each cage, implying that on the cage centered around S$_{iso}$ there exists a smooth PES for Li$^+$ ions, which is a characteristic indicator for frustration phenomenon in fast ion conductors.

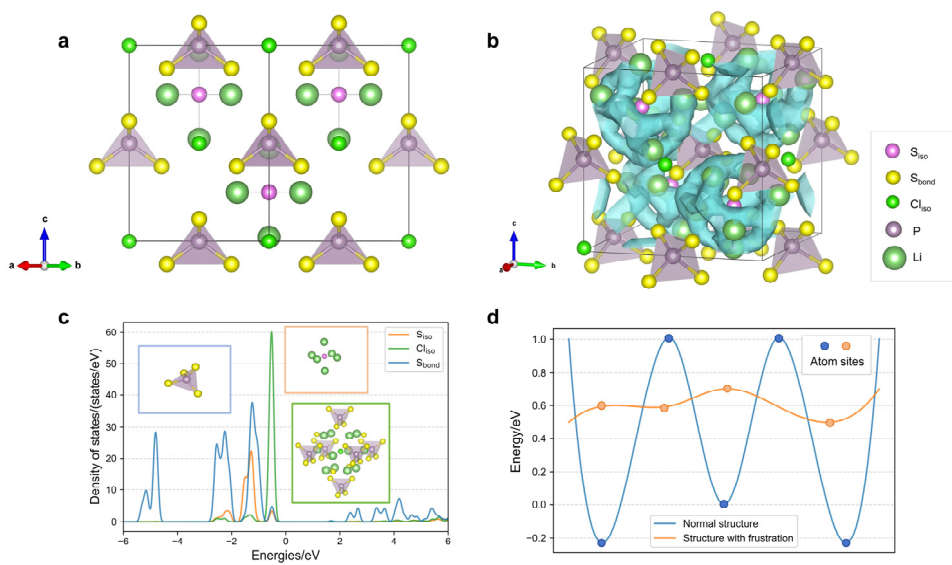

Figure 1: Isolated anions in argyrodites and the resulting frustration.

**a** Crystal structure of $Li_6PS_5Cl$. **b** $Li^+$ diffusion pathways in $Li_6PS_5Cl$ calculated through AIMD, the areas within the blue surfaces are positions with high probability density of $Li^+$ ions. **c** The local environment of $S_{iso}$ (the inset in the orange box), $Cl_{iso}$ (the inset in the green box) and $S_{bond}$ (the inset in the blue box) and the PDOS of $S_{iso}$, $Cl_{iso}$ and $S_{bond}$ in $Li_6PS_5Cl$. **d** A schematic diagram of energy landscape for $Li^+$ migration in normal structure and structure with frustration.

The frustration in superionic conductors, acting as one of the effective migration mechanisms for high conductivity of $Li^+$ ions, might be caused by local distortion or by partial occupation. For example, Stefano et al.[25] showed that the disorder distortion of $[PS_4]^{3-}$ groups in $LiTi_2(PS_4)_3$ triggers the frustration phenomenon by offering no energetically stable tetrahedral or octahedral positions for the $Li^+$ ions in the structure. Wang et al.[26] proposed the Density Of Atomistic States (DOAS) to quantitatively characterize the degree of disordering and elucidated how the frustration enhances diffusion. The distribution of DOAS also reflects the similarity of local environment for $Li^+$ ions on the smooth PES. Besides DOAS, the distance of atomic descriptors between positions in the structure can also tell the degree of the similarity in the local chemical environments and local configuration energy. By computing the Euclidean distances of Atom-Centered Symmetry Functions (ACSFs) descriptor[27] between positions in the lattice of $Li_6PS_5Cl$ (Supplementary Figure S1 and Supplementary Note 1), the positions similar to $Li^+$ lattice sites were picked out and the results show that they are all distributed on Li cages surrounding $S_{iso}$, indicating the presence of frustration around $S_{iso}$. In general, the frustrated system has a large number of degenerated states with similar energies. The exceptionally smooth energy landscape, comparing with normal structures with specific stable sites for $Li^+$ ions (Figure 1d), ensures that the system can achieve lower energy barrier transitions between degenerated states.

Although the role of frustration phenomenon in ion transport has been recognized, there still exist mysteries such as the features of structures with frustration, the criteria for determining if a frustration appears and the design method for frustrated system. In this work, by thoroughly investigating the ion migration behaviors in a series of structures of the prototype compound $Li_8SiSe_6$, we found that isolated anions can be one of the effective feature environments for introducing frustration in crystal structures. The definition of isolated anions, as well as how their local symmetries correlate with the appearance of frustration, and how the arrangement and type of anions affect the ionic transport are discussed. Using isolated anions as the structural feature, we screened compounds containing this feature from the crystal structure database, and further AIMD studies confirms the fast ion migration behavior in these compounds, proving the function of isolated anions in creating high ionic conductivity in these structures.

There are many problems worth exploring regarding the isolated anions. Most argyrodites show different phases at the high- and low-temperature regions.[28] Generally, the high-temperature phase has higher symmetry as well as higher ionic conductivity. Also, the doping of halogen element (Cl, Br) helps to stabilize the high-temperature phase at room temperature,[29] for example, $Li_6PS_5Cl$, but the $Cl_{iso}$ or $Br_{iso}$ anions do not generate Li cages around them as $S_{iso}$ do. Therefore, the symmetry of the structures, the arrangements and the types of isolated anions are all important factors related to frustration and $Li^+$ transport. The reasons for choosing $Li_8SiSe_6$ as prototype structures are as follows. Firstly, $Li_8SiSe_6$ structures contain both isolated Se ($Se_{iso}$) and bonded Se ($Se_{bond}$), which is necessary for our research. Secondly, the diverse and abundant phase structures of $Li_8SiSe_6$ can be derived from other existing argyrodite materials through element substitution (Table 1), and these phases have varying levels of ion transport capabilities, providing possibility to construct relationship between various structural variables and ionic conductivity. Thirdly, phases of $Li_8SiSe_6$ show relatively high ionic conductivities at 300 K or 400 K compared with other argyrodites, eliminating the necessity for extrapolating room temperature conductivity by high-temperature AIMD simulations, hence the influence of phase transition caused by temperature change can be removed. Finally, in $Li_8SiSe_6$, there is only one type of anion element, Se, which reduces the influence of doping halogen element such as Cl or Br, making less variables and making the initial research simpler and more intuitive. Halogen elements can also be introduced to replace part of isolated Se sites for comparative research to clarify the effects of anion types. Overall, six different space group structures of $Li_8SiSe_6$ ($Li_8SiSe_6\_F\bar{4}3m$, $Li_8SiSe_6\_Pna2_1$, $Li_8SiSe_6\_Pmn2_1$, $Li_8SiSe_6\_P6_3cm$, $Li_8SiSe_6\_Cc$ and $Li_8SiSe_6\_hcp$) as well as $Li_7SiSe_5Cl$ with $F\bar{4}3m$ space group were considered in our research system (Table 1). The feature of isolated anions is also applied to look for new structural frames potentially used as fast ion conductors. The obtained conductors screened from crystal structure database include $Li_6NI_3$, $Li_7N_2I$, $Li_5CrCl_8$, $Li_6VCl_8$, $Li_5SbS_3I_2$ and $Li_8TiS_6$. The structural characteristics are discussed in detail and the thermostability and ion migration properties are investigated by DFT structural relaxation and AIMD simulations for each candidate structure, ensuring that the feature of isolated anions is an effective way to screen and design fast ion conductors with frustration mechanism.

Table 1: Basic information of the analyzed structures. (Among them, $Li_8SiSe_6\_P6_3cm$ and $Li_8SiSe_6\_hcp$ are both $P6_3cm$ space group but their arrangements of isolated anions are totally different, so one of them is called $Li_8SiSe_6\_hcp$ to discriminate.)

| Structure | Origination | $E_{hull}$/ (meV/atom) | Diffusivity/ (cm$^2$/s) | Ionic conductivity/ (mS/cm) |
|---|---|---|---|---|
| $Li_8SiSe_6\_F\bar{4}3m$ | $Ag_8SiS_6\_F\bar{4}3m$(icsd_605720) | 181 | $2.73 \times 10^{-6}$ (300K) | 417.2 (300K) |
| $Li_8SiSe_6\_Pna2_1$ | $Ag_8SiS_6\_Pna2_1$ (icsd_001054) | 7 | $2.65 \times 10^{-6}$ (400K) | 332.2 (400K) |
| $Li_8SiSe_6\_Pmn2_1$ | $Cu_8SiSe_6\_Pmn2_1$ (icsd_089451) | 23 | $1.57 \times 10^{-6}$ (300K) | 261.3 (300K) |
| $Li_8SiSe_6\_P6_3cm$ | $Cu_8GeSe_6\_P6_3cm$ (mp-570393) | 27 | $1.38 \times 10^{-6}$ (300K) | 226.2 (300K) |
| $Li_8SiSe_6\_Cc$ | $Cu_8GeSe_6\_Cc$ (mp-1225860) | 11 | $1.26 \times 10^{-6}$ (400K) | 158.5 (400K) |
| $Li_8SiSe_6\_hcp$ | $Li_8SiO_6\_hcp$ (mp-28549) | 0 | $7.11 \times 10^{-8}$ (400K) | - (400K) |
| $Li_7SiSe_5Cl$ | $Li_7Zn_{0.5}SiS_6$[43] | 44 | $2.57 \times 10^{-6}$ (300K) | 149.3 (300K) |

# Results

**Definition and characteristics of isolated anions**

Anions those exist independently in structures and do not form local structures such as tetrahedral or octahedral coordinating with immobile non-lithium cations, are defined as isolated anions in our study. As shown in Figure 2a, the $Se^{2-}$ anions are classified into $Se_{bond}$ and $Se_{iso}$ according to whether it appears at the corner of $[SiSe_4]^{4-}$ tetrahdron or as the isolated one only surrounded by $Li^+$ ions. Besides the different bonding characteristics, these two types of Se anions also exhibit clear differences in their electronic structures. The PDOS and vacancy formation energy for $Se_{bond}$ and $Se_{iso}$ in $Li_8SiSe_6$ argyrodites are shown in Figure 2b, 2c and Supplementary Figure S2. In $Li_8SiSe_6\_F\bar{4}3m$, the PDOS of $Se_{iso}$ overlaps with part of Li's PDOS and no superimposing PDOS between $Se_{iso}$ and Si, indicating bonds with only Li, and the PDOS of $Se_{iso1}$ and $Se_{iso2}$ are distributed in different energy range due to their distinct bonding numbers to Li as shown in Figure 2a. However, the PDOS of $Se_{bond}$ overlaps with PDOS of both Si and Li, implying bonds with both Si and Li. The vacancy formation energy of $Se_{iso}$ is larger than that of $Se_{bond}$ which illustrates that $Se_{iso}$ is more stable in structures. The unique local environment of isolated anions makes them show strong influence on $Li^+$ ion transport. So next, after evaluating the ionic conductivity in each structure, we will figure out potential influencing factors one by one and study how the isolated anions affect $Li^+$ ion transport and their correlation with frustration mechanism.

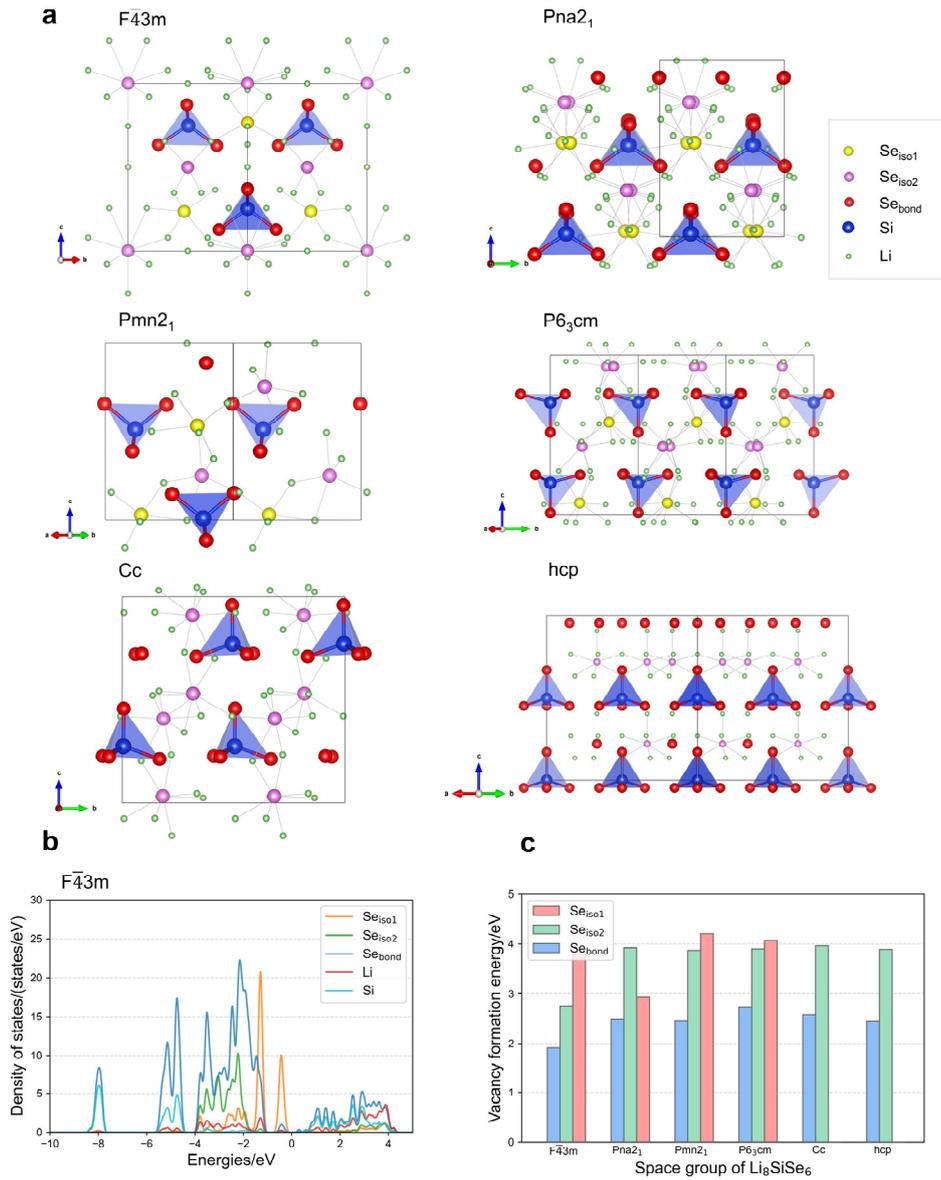

Figure 2: The characteristics of Se$_{iso}$ and Se$_{bond}$ in Li$_8$SiSe$_6$ structures.

**a** The structures of Li$_8$SiSe$_6$. **b** PDOS in Li$_8$SiSe$_6$_F$\bar{4}$3m. **c** Vacancy formation energy of Se$_{iso}$ and Se$_{bond}$ in Li$_8$SiSe$_6$ structures. In Li$_8$SiSe$_6$_F$\bar{4}$3m, Li$_8$SiSe$_6$_Pna2$_1$, Li$_8$SiSe$_6$_Pmn2$_1$, and Li$_8$SiSe$_6$_P6$_3$cm, Se$_{iso}$ can be divided into Se$_{iso1}$ and Se$_{iso2}$ due to different Wyckoff positions. The calculation process of vacancy formation energy is described in Supplementary Note 2.

**Li ion transport properties in Li$_8$SiSe$_6$ and Li$_7$SiSe$_5$Cl structures**

The Li$^+$ ion diffusivity and ionic conductivity are evaluated for each Li$_8$SiSe$_6$ structure by fitting the MSD curves statistically extracted from AIMD simulations at 300K or 400K (Table 1 and Supplementary Figure S3). All six structures exhibit a cage-like transport mechanism surrounding the Se$_{iso}$ (Supplementary Figure S4). Li$_7$SiSe$_5$Cl's MSDs, diffusivity and conductivity at 300K are also obtained (Supplementary Figure S5, Table 1). As Table 1 shows, argyrodites containing Se element can already achieve very high Li$^+$ conductivity without the existence of halogen element. For Li$_8$SiSe$_6$ in the space group of F$\bar{4}$3m, Pmn2$_1$ and P6$_3$cm, ion migration events can even be observed at 300 K. However, both Li$_8$SiSe$_6$_hcp and Li$_8$SiSe$_6$_Cc structures exhibit no Li$^+$ motion at 300 K, and the MSD of Li$^+$ ions in Li$_8$SiSe$_6$_Pna2$_1$ is too small to estimate the diffusivity and ionic conductivity. Perform AIMD simulations at 400 K for the last three compounds, and the kinetic properties can be obtained for Li$_8$SiSe$_6$_Pna2$_1$ and Li$_8$SiSe$_6$_Cc, while Li$_8$SiSe$_6$_hcp still shows no hopping events at 400K. To sum up, the ionic transport ability of the structures of Li$_8$SiSe$_6$ is ranked as: Li$_8$SiSe$_6$_F$\bar{4}$3m > Li$_8$SiSe$_6$_Pmn2$_1$ > Li$_8$SiSe$_6$_P6$_3$cm > Li$_7$SiSe$_5$Cl > Li$_8$SiSe$_6$_Pna2$_1$ > Li$_8$SiSe$_6$_Cc > Li$_8$SiSe$_6$_hcp. Comparing the spatial positions of cages in Li$_8$SiSe$_6$_F$\bar{4}$3m and Li$_7$SiSe$_5$Cl, it can be found that the neighboring cages in Li$_8$SiSe$_6$_F$\bar{4}$3m share a common face, while the cages in Li$_7$SiSe$_5$Cl are far apart resulting in distinguishable intra- and inter-cage structures (Supplementary Figure S6). In addition, the face-sharing cages also exist in Li$_8$SiSe$_6$_Pna2$_1$, Li$_8$SiSe$_6$_Pmn2$_1$, Li$_8$SiSe$_6$_P6$_3$cm and Li$_8$SiSe$_6$_Cc, but Li$_8$SiSe$_6$_hcp shows similar case to Li$_7$SiSe$_5$Cl in which cages are apart from each other. We can deduce that different local environments and spatial arrangements of isolated anions have an impact on the formation of cages as well as the ion migration. Therefore, we will systematically discuss the influence of various factors, such as symmetry, arrangement, and elemental type, on the ion migration properties in detail.

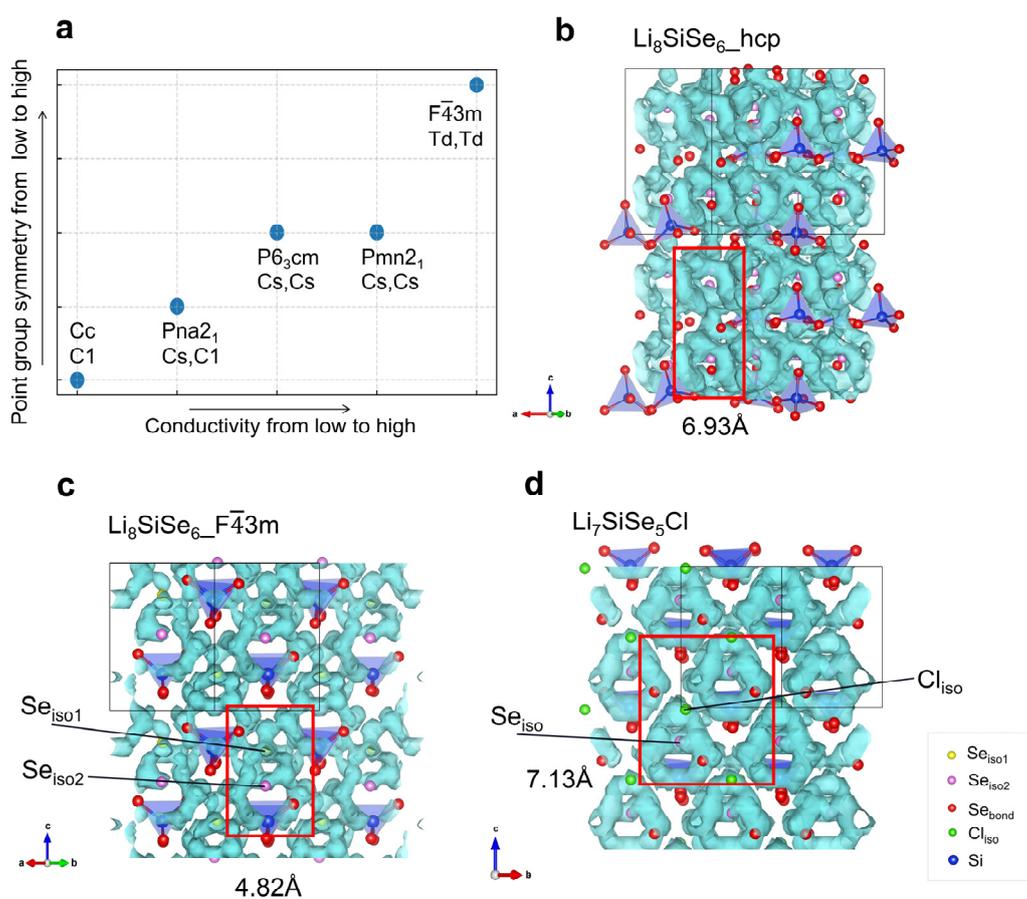

Figure 3: The effects of isolated anions on Li$^+$ ion transport.

**a** A schematic diagram of the relationship between Se$_{iso}$ local structure's point group symmetry and Li$^+$ conductivity. **b-d** Li$^+$ diffusion pathways obtained from AIMD simulations in Li$_8$SiSe$_6$_hcp, Li$_8$SiSe$_6$_F$\bar{4}$3m and Li$_7$SiSe$_5$Cl. The values are distances between adjacent Se$_{iso}$ in red boxes. The Li$^+$ ions in **b-d** are hidden for better observation of the pathways.

**The effect of the symmetry of isolated anions' local structures on intra-cage transport**

The local structural features for Se$_{iso}$ anions in each compound are analyzed to build the relation to the kinetic properties. For a specific Se$_{iso}$ anion, the point group of the adjacent [SiSe$_4$]$^{4-}$ tetrahedra and other Se$_{iso}$ is adopted to describe its local symmetry. The symmetry of local environment can represent the symmetry of Li cages' PES around $Se_{iso}$. The intuitive understanding is that a highly symmetric local environment will lead to the PES around $Se_{iso}$ having high symmetry, which is more conducive to the formation of the cage-like pathway for lithium ions.

For the five structures with face-sharing $Li^+$ migration cages as mentioned above, the inter-cage transport can be accomplished through these sharing faces, and thus they are a suitable set of systems for studying the influence of local symmetry on intra-cage transport. The point group of local environments for $Se_{iso}$ in these five structures are determined and shown in Figure 3a and Supplementary Figure S7. Three types of point group environment are found, including Td, Cs, and C1. It is found that structures with higher local symmetry also exhibit higher ionic conductivity, which occurs through the hopping of $Li^+$ ions within the cage surfaces, known as intra-cage pathways around $Se_{iso}$. By controlling synthesis conditions or substitution, the lattice space group can be adjusted by forming various phases and the symmetry of local structures can be controlled to improve ion transport performance in ion conductors[30]. However, for the remaining structure, $Li_8SiSe_6\_hcp$, the point group of local environment for $Se_{iso}$ is C3. The high symmetry maintains the high degree of frustration in the region of cage surfaces around $Se_{iso}$, but the distance between cages restricts the ion migration within the entire crystal framework. As a result, hcp phase shows relatively low ion conductivity, thus we will explore the conditions for reducing the energy barrier of inter-cage transport from the perspective of isolated anion arrangements.

**The effect of the arrangement of isolated anions on inter-cage transport**

Since the isolated anions act as the centers of the cage pathways, the arrangement of them will determine the position of the cages and affect the ion transfer among these cages. When the isolated anions are closely packed in a structure, coplanar transport can be easily formed to connect pathways among different cages, allowing for inter-cage transfer to occur directly, as found in the five structures in Figure 3a. The distances between neighboring cage centers are 4.82 Å for $Li_8SiSe_6\_F\bar{4}3m$, 4.45 Å for $Li_8SiSe_6\_Pna2_1$, 4.49 Å for $Li_8SiSe_6\_Pmn2_1$, 4.27 Å for $Li_8SiSe_6\_P6_3cm$ and 4.15 Å for $Li_8SiSe_6\_Cc$. While for structure $Li_8SiSe_6\_hcp$ (Figure 3b), the distance between adjacent $Se_{iso}$ anions is as large as 6.93 Å, where the cages can't be connected directly, thereby the inter-cage transport becomes a major limiting factor for Li motion and results in the relatively low $Li^+$ ion conductivity. According to the atomic radius of Li and Se's, the maximum $Se_{iso}$ spacing of 5.22 Å is estimated as the critical distance to achieve coplanar cage transport. On the other hand, when the distance between $Se_{iso}$ and other non-Li ions is too small to permit $Li^+$ pass, the cages can't form successfully. Therefore, high conductivity can be obtained by constructing coplanar Li ion cages through shortening the distance of adjacent isolated anions to below critical distance while maintaining the geometric size of the $Li^+$ migration tunnels.

**The effect of isolated anions' types on Li cage connectivity**

When $Li_8SiSe_6\_F\bar{4}3m$ is compared with $Li_7SiSe_5Cl$, it is found that even though the Cl-doped $F\bar{4}3m$ structure has much lower $E_{hull}$ than the original one (Table 1), no $Li^+$ cage pathways around $Cl_{iso}$ can be found and only the intra-cage transport around $Se_{iso}$ is maintained. The disappearance of cages around $Cl_{iso}$ sites leads to the increase of the distance between isolated anions from 4.82 Å to 7.13 Å, and the face-sharing inter-cage transport are damaged (Figure 3d), thus the doped structure $Li_7SiSe_5Cl$ has a lower Li MSD and conductivity at 300K than $Li_8SiSe_6\_F\bar{4}3m$ (Table 1 and Supplementary Figure S3, S5). Ouyang et al.[31] explained the phenomenon in Na argyrodite by analyzing the Na-Na intracluster and intercluster distances of every structure. They pointed out that halogen doping can change these distances in structures and the distance variation is created by the size contrast between the halogen and the sulfur. Here we illustrate this phenomenon and characterize it quantitatively from the perspective of interactions between Li$^+$ and isolated anions. For the cases with two different types of anionic elements, Li$^+$ ions may move closer to anions with stronger interactions and form cage-like pathways around them, while the probability of lithium ions appearing around anions with weaker interactions decreases, causing the disappearance of the cages. For simplicity, the Ewald energy is used to evaluate the strength of electrostatic interactions between lithium ions and isolated anions. In Li$_8$SiSe$_6$_F$\bar{4}$3m, the two sites of Se$_{iso}$ have similar Ewald energy (-18.06eV and -14.74eV). However, in Li$_7$SiSe$_5$Cl, the results reveal that Cl$_{iso}$ (-4.55eV) has much higher Ewald energy than Se$_{iso}$ (-17.41eV) and Li$^+$ ions tend to distribute around Se$_{iso}$ rather than Cl$_{iso}$. Additionally, here is another typical example, two different isolated anions, I$_{iso}$ and S$_{iso}$, exist in solid electrolyte Li$_7$P$_2$S$_8$I[32, 33], in which the Ewald energy of I$_{iso}$ (-2.61eV) is much higher than S$_{iso}$ (-15.96eV), leading to the formation of Li$^+$ cage pathways only surround S$_{iso}$ according to the Li$^+$ trajectories revealed by AIMD at 600K (Supplementary Figure S8). The Ewald energy of isolated anions can be used as a practical parameter to determine in which part the cage pathways can be formed if multiple types of isolated anions exist in the structure. It can be served as the indicator to design which element should be doped to adjust the connectivity of the cages by modulating the interactions between Li$^+$ ions and different types of isolated anions.

According to above characteristics summarized from Li$_8$SiSe$_6$, we deduce that the existence of isolated anions in structures is one of the ways to trigger cage transport mechanism of Li$^+$ ions, the part of face sharing between neighbouring cages can further reduce the barrier of inter-cage transport and result in high ionic conductivity. Six structures of Li$_8$SiSe$_6$ with various space group and Li$_7$SiSe$_5$Cl_F$\bar{4}$3m are taken as examples to analyse the factors influencing cage transport. First, the symmetry of isolated anions' local structures affects intra-cage transport of Li$^+$ ions, and the high symmetry enables the presence of more similar energy sites of Li$^+$ around isolated anions and thus produce the frustration phenomenon. Second, the arrangement of isolated anions plays an important role in the inter-cage transport of Li$^+$ ions. The densely packed isolated anions are more likely to produce cages with sharing faces, thus eliminate the limiting step of the inter-cage transport. Finally, when multiple types of isolated anions exist, the competition of the interaction between Li$^+$ and different isolated anions affects the connectivity of the cages. The Ewald energy can help us to judge around which anions the cage pathways may be connected. The transport mechanism related to the isolated anions can be applied to discover new fast ion conductors. By conducting the high-throughput screening from crystal structure databases based on the features of isolated anions, we have identified several types of fast ion conductors and the studies on their ion migration properties are carried out.

**New conductors with isolated anions obtained by high-throughput screening**

Since argyrodites with Li cages around isolated anions behave well in ionic transport, the high-throughput screening[34] is performed to look for the crystal structure frameworks those share the isolated anion feature.[35] The structures in the Inorganic Crystal Structure Database (ICSD)[36] and Materials Project (MP)[37, 38] database are considered. Four main criterion of the screening process is illustrated in Figure 4a. Only the Li-contained compounds are considered. As a verification of the findings on isolated anions, we limit the screening process in ternary and quaternary compounds. Structures composed of more elements can be investigated using the same methodology. The $E_{\text{hull}}$ values of the candidates are less than 15 meV/atom to keep the possibility for synthesis. Among them, the structures with isolated anions are recognized by bonding analysis and picked up for kinetic property simulations. Through these criterion, four types of structures with different conducting related isolated anions are obtained. Each type of materials contains one of the isolated anions related to ion transport: $N^{3-}$, $Cl^-$, $I^-$ or $S^{2-}$, and has its own structural characteristics. The point group of isolated anions's local structure and the adjacent distances between them have been extracted for these structures. The former is identified to evaluate the intra-cage transport performance of the structure, and the latter generally relates to the inter-cage transport and acts as the limiting factor for ion transport. The information of isolated anions in conductors, including their element type, their local structures' point group as well as distances between neighboring isolated anions are illustrated in Table 2. And the conductors' structural and ion conducting properties obtained by AIMD are shown in Figure 4 and Supplementary Figure S9.

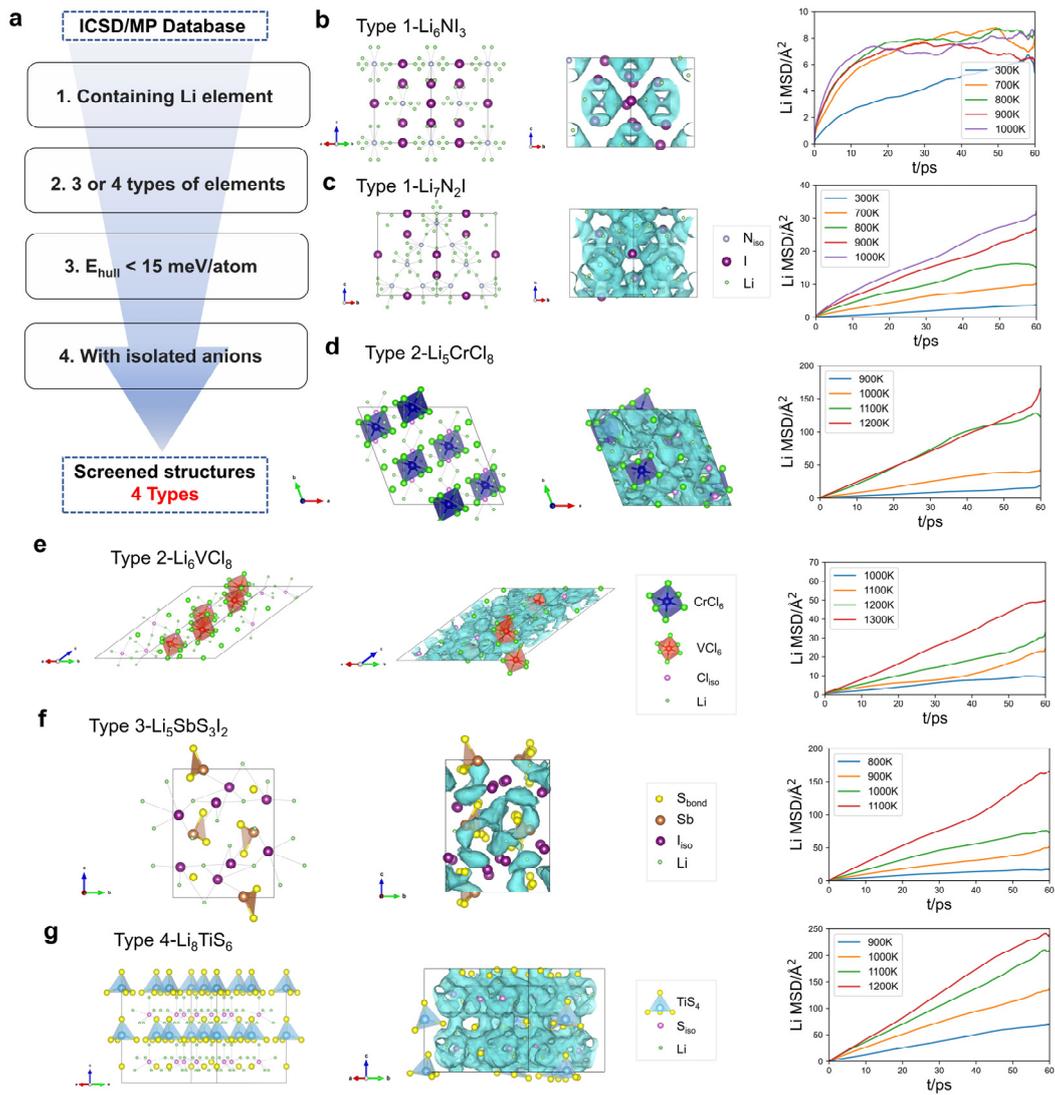

Figure 4: Discovery of new conductors with isolated anions by high-throughput screening.

**a** High-throughput screening process looking for Li$^+$ ion conductors with structural features of isolated anions. **b-g** Structures, Li$^+$ diffusion pathways and MSDs of new conductors calculated by AIMD.

The first type relates to the N$^{3-}$ anions. Li$_6$NI$_3$ and Li$_7$N$_2$I, as shown in Figure 4b and 4c, have similar structural frameworks possessing an alternating distribution of isolated I$^-$ and N$^{3-}$ atoms. We notice that Li$_7$N$_2$I with 0.5% carbon nanotube compounds are recently discovered experimentally and has a high ionic conductivity of $3.1 \times 10^{-4}$ S/cm[39], and many studies have also reported Li-N-X (X = I, Cl or S) electrolytes with isolated $N^{3-}$ ions[40-42], ensuring the effectiveness of the screening. In these structures, $Li^+$ ions are primarily distributed around $N^{3-}$ for the reason that the Ewald energy of $Li^+$-$N^{3-}$ interaction (-33.27 eV in $Li_6NI_3$ and -48.88 eV in $Li_7N_2I$) is lower than that of $Li^+$-$I^-$ interaction (-5.25 eV in $Li_6NI_3$ and -4.22 eV in $Li_7N_2I$). In $Li_7N_2I$, the point group symmetry of the isolated $N^{3-}$ ions' local structure is Cs, which is lower than that of $Li_6NI_3$, Oh, but the distance between adjacent isolated $N^{3-}$ ions (3.85 Å) is quite smaller than that in $Li_6NI_3$ (6.32 Å). It can be inferred that compared with $Li_6NI_3$, $Li_7N_2I$ has a worse intra-cage transport of $Li^+$ ions, but a better inter-cage transport. The AIMD results further confirm the above inference. First of all, in both structures, the cage transport exists around $N^{3-}$ ions even at the AIMD simulations of 300 K, indicating that the frustration phenomenon is successfully formed because of the isolated anions. Only intra-cage transport of $Li^+$ ions is found in $Li_6NI_3$ according to the fact that its MSDs at different temperatures (300K, 700K, 800K, 900K and 1000K) have almost the same upper limit, and there is no connection between neighboring cages observed according to the trajectories of $Li^+$ ions. However, for $Li_7N_2I$, it has lower MSD at 300K (mainly low-barrier intra-cage transport) but much larger MSDs at higher temperatures (achieving inter-cage transport). The worse intra-cage transport in $Li_7N_2I$ is due to the lower symmetry of isolated $N^{3-}$'s local structure and the easier inter-cage transport is attributed to the reduction of distances of adjacent isolated $N^{3-}$ ions compared with $Li_6NI_3$. The face sharing Li cages in $Li_7N_2I$ is also manifested by $Li^+$ diffusion pathway. In addition to $Li_6NI_3$ and $Li_7N_2I$, the other screened structures containing isolated $N^{3-}$ anions are listed in Supplementary Table S1.

The second type is $Li_5CrCl_8$ and $Li_6VCl_8$ (Figure 4d, 4e) containing isolated $Cl^-$ anions. In $Li_6VCl_8$, isolated $Cl^-$ anions' local structure point group symmetry is Td, higher than that of C2v in $Li_5CrCl_8$, but the isolated $Cl^-$ anions are more loosely packed than in $Li_5CrCl_8$, indicating better intra-cage transport and worse inter-cage transport compared with $Li_5CrCl_8$, which is similar to the case in $Li_6NI_3$ and $Li_7N_2I$. But unfortunately, the obvious $Li^+$ migration in these structure can only occur in AIMD at temperatures above 900 K. Although higher ionic conductivity is predicted in $Li_5CrCl_8$ than in $Li_6VCl_8$, the high barriers of the inter-cage transport act as the control factor of ion transport in both compounds. Furthermore, the other screened compound $Li_6NiCl_8$ is also listed in Supplementary Table S1.

The third type is $Li_5SbS_3I_2$ (Figure 4f) with isolated $I^-$ anions and it shows a relatively high conductivity. In this structure, the symmetry of $I^-$ ions' local environment is Cs and the distance between $I_{iso}$ is 4.02 Å and 4.36 Å. Although the structural characteristics meet the criteria, the complete cage-like pathways are not observed in the trajectories of AIMD. Upon closer examination of its crystal structure, it can be observed that the presence of $[SbS_3]^{3-}$ groups disrupts the formation of cage channels. There is no tetrahedron formed by $Sb^{3+}$ and $S^{2-}$ and the distance between $Sb^{3+}$ and $I^-$ is only 3.70 Å, so the existence of $Sb^{3+}$ around isolated $I^-$ sets up obstacles for creating complete cage-like pathways. But the low-barrier inter-cage transport, arising from the close proximity of isolated $I^-$ anions, provides a favorable pathway for the $Li^+$ ion migration. Therefore, the element types of the surrounding ions of isolated anions can also affect the appearance of the cages. It also tells us additional information that the appearance of the pathways around the isolated anions can be adjusted through changing the position and environment of non-Li cations.

The fourth type is $Li_8TiS_6$ (Figure 4g), it has the same structure as $Li_8SiSe_6\_hcp$. It does not exhibit $Li^+$ migration in AIMD simulations at 300 K, and the ion conductivity at 300K is obtained through extrapolation of ionic conductivities at high temperatures.

Table 2: Basic information of the new conductors.

| Structure | Origination | Isolated anions' type | $E_{hull}$/ (meV/atom) | Diffusivity/ (cm$^2$/s) | Ionic conductivity/ (mS/cm) | Isolated anions' local structure's point group | Distance between isolated anions/Å | Anion lattice type |
|---|---|---|---|---|---|---|---|---|
| Li$_6$NI$_3$ | icsd-083380 | N | - | 1.74 × 10$^{-6}$ (300K) | 361.5 (300K) | Oh | 6.32 | fcc |
| Li$_7$N$_2$I | icsd-085713 | N | - | 1.05 × 10$^{-6}$ (300K) 8.63 × 10$^{-6}$ (1000K) | 6.8 (300K); 801.8 (1000K) | Cs | 3.85 | fcc |
| Li$_5$CrCl$_8$ | mp-23361 | Cl | 11 | 1.42 × 10$^{-14}$ (300K) 1.29 × 10$^{-5}$ (1000K) | 1.61 * 10$^{-6}$ (300K) 440.8 (1000K) | C2v | 3.66; 5.06 | fcc |
| Li$_6$VCl$_8$ | mp-29250 | Cl | 13 | 4.17 × 10$^{-13}$ (300K) 2.90 × 10$^{-6}$ (1000K) | 5.64 * 10$^{-5}$ (300K) 117.8 (1000K) | Td | 5.16 | fcc |
| Li$_5$SbS$_3$I$_2$ | mp-559814 | I | 0 | 1.01 × 10$^{-11}$ (300K) 2.24 × 10$^{-5}$ (1000K) | 0.001 (300K) 772.3 (1000K) | Cs | 4.02; 4.36 | sc |
| Li$_8$TiS$_6$ | mp-753546 | S | 0 | 8.90 × 10$^{-10}$ (300K) 3.84 × 10$^{-5}$ (1000K) | 0.2 (300K) 2075.8 (1000K) | C3 | 3.97; 6.69 | hcp |

The successful detection of the four types of Li conductors with isolated anions of N$^{3-}$, Cl$^-$, I$^-$, and S$^{2-}$ confirms the feasibility of utilizing isolated anions as a screening feature. In addition, the anionic lattices of the above new conductors are not bcc but fcc, hcp or simple cubic (sc) structures as Table 2 shows, it implies that isolated anions can be introduced to structures with non-bcc anionic lattices to improve ion migration. And the influencing factors derived from argyrodite Li$_8$SiSe$_6$ are universally applicable to the analysis of pathways in such structures.

# Discussion

In this study, we propose that isolated anions can serve as a structural feature for the formation of cage-like transport pathways, through which frustration phenomenon always appears and high ionic conductivity will be expected. Therefore, introducing and appropriately increasing the proportion of isolated anions in structures can be one of the methods to improve the ionic conductivity. Compared with bonded anions, isolated anions always show PDOS overlapping only with electronic states of $Li^+$ ions and higher vacancy formation energy, which reveal their unique bonding and high stability.

By researching six $Li_8SiSe_6$ structures with different space group and $Li_7SiSe_5Cl$ with $F\bar{4}3m$ group, three factors related to isolated anions those influence $Li^+$ ion transport are derived. To begin with, the symmetry of isolated anions' local structure decide intra-cage transport and higher symmetry brings better intra-cage transport. Therefore, frustration phenomena can be created by introducing isolated anions and constructing their local environments with high symmetry. Following this principle, promising argyrodites ionic conductors can be predicted relying solely on structural regulation and this breaks the traditional idea of increasing argyrodites' conductivity by stabilizing the cubic phase at room temperature through doping halogen element. Besides, as for inter-cage transport, the arrangement of isolated anions become the most important factor because it is directly related to whether the cages are face-sharing connected or not, and the densely packed anions produce better inter-cage transport. This can explain the reason why some structures contain cage transport channels but can't reach high conductivity. It also means that promoting the generation of face sharing cages through structural modification should be a promising strategy to improve ionic conductivity. Furthermore, when a structure owns several types of isolated anions, the interaction between different types of isolated anions and Li$^+$ ions varies, thus affecting the connectivity inside the cage and the Ewald energy can help us quantify the impact and provide indications for element doping. All in all, the characteristics of isolated anions play a significant role in the discovery and design of ion conductors with frustration phenomena.

Additionally, by high-throughput method, four types of new Li$^+$ ion conductors which contain isolated anions are identified. The ion transport behaviors in these conductors are demonstrated and analyzed by AIMD calculations, indicating that the principles derived from argyrodite Li$_8$SiSe$_6$ and Li$_7$SiSe$_5$Cl are suited generally for structures with feature of isolated anions. Our conclusions and findings of isolated anions can be utilized for explaining a wide range of phenomena in structures with cage transport mechanism and can guide the design and improvement of these structures to reach higher ion conductivity. However, in what kind of structures isolated anions can be introduced and how the stability can be maintained are still open questions and further exploration is required.

## Methods

**Materials**

Li$_6$PS$_5$Cl structure was obtained in Materials Project database (mp-985592). Six different structures of argyrodite Li$_8$SiSe$_6$ and the structure of Li$_7$SiSe$_5$Cl were obtained by atom substitution of other argyrodites in Materials Project database, ICSD database and the cif file of Li$_7$Zn$_{0.5}$SiS$_6$[43] (Table 1). Li$_7$P$_2$S$_8$I structure was obtained in references 32 and 33. For high-throughput search, the new conductors were screened from both Materials Project database and ICSD database. The structures were visualized by the VESTA software package[44].

**Density functional theory computation of structural relaxation and static calculations**

We carried out all the relaxation calculations in primitive cells (Supplementary Table S2 lists the total number of atoms for each structure) using Vienna Ab initio Simulation Package (VASP)[45] based on density functional theory (DFT) using Perdew–Burke–Ernzerhof (PBE)[46] generalized gradient approximation (GGA) described by the projector-augmented-wave (PAW) approach. The cutoffs for the wave function and density are 520eV and 780eV, respectively. Both ions and cells were relaxed in the optimization, with the energy and force convergence criterion of $10^{-5}$ eV and 0.01 eV/Å, respectively. Spin-polarized DFT was used to relax $Li_5CrCl_8$ and $Li_6VCl_8$. The $E_{hull}$ used to assess the thermodynamic stability was calculated using phase diagrams in the Materials Project (MP) database[47,48]. The point group of isolated anions' local structure was determined by pymatgen.symmetry.analyzer.PointGroupAnalyzer Python package.[49] The Ewald energy used to determine the interactions between anions and $Li^+$ ion was calculated by using the pymatgen.analysis.ewald Python packages.

**Ab initio molecular dynamics simulations**

We performed AIMD simulations to study ionic diffusion in supercell models (Supplementary Table S2 lists the total number of atoms for each structure) with lattice parameters around 10 Å, using nonspin-polarized DFT calculations with a Γ-centered k-point. Spin-polarized DFT was used to simulate $Li_5CrCl_8$ and $Li_6VCl_8$. The AIMD run was carried out with a Nose thermostat[50] for 70,000 steps, with a time step of 1 fs. We picked different temperatures to simulate different materials. The first 10 ps is used for structural equilibrium and the last 60 ps is used for kinetic property analysis. The results of them were used to investigate the diffusivity, ionic conductivity as well as diffusion pathways. The ionic conductivity was calculated following the established method[51]. We used positions of high probability density of Li ions calculated as the fraction of time that each spatial location is occupied to describe the diffusion pathways of Li[52]. We performed the diffusion analysis by using pymatgen and the pymatgen.analysis.diffusion.analyzer Python packages[53].

# Data availability

All raw data generated during the study are available from the corresponding authors upon request.

# Acknowledgements

This work was supported by funding from the National Natural Science Foundation of China (grants no. 52172258, 52022106), the Strategic Priority Research Program of Chinese Academy of Sciences (grant no. XDB0500200). The numerical calculations in this study were carried out on both the ORISE Supercomputer, and the National Supercomputer Center in Tianjin.

# Competing interests

The authors declare no competing interests.

# Supplementary Materials for "Isolated anions induced high ionic conductivity"

Qifan Yang[1,2], Jing Xu[1,3], Yuqi Wang[1,3], Xiao Fu[1,2], Ruijuan Xiao[1,2,3]* and Hong Li[1,2,3]*

[1] Beijing National Laboratory for Condensed Matter Physics, Institute of Physics, Chinese Academy of Sciences, Beijing 100190, China.

[2] Center of Materials Science and Optoelectronics Engineering, University of Chinese Academy of Sciences, Beijing 100049, China

[3] School of Physical Sciences, University of Chinese Academy of Sciences, Beijing 100049, China

E-mail: rjxiao@iphy.ac.cn, hli@iphy.ac.cn

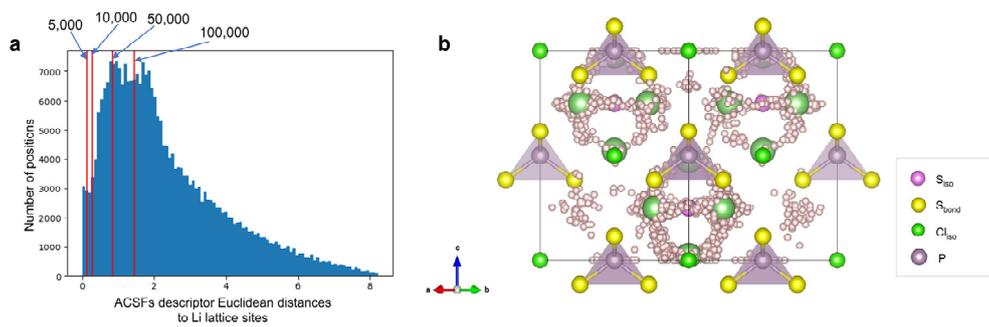

**Supplementary Figure S1. The ACSFs descriptor analysis of Li$_6$PS$_5$Cl.**
**a** The number of positions with different ACSFs descriptor Euclidean distances to Li lattice sites in Li$_6$PS$_5$Cl, and the values represent the total numbers of positions from origin to the red vertical line it pointed to. **b** Randomly select 1,500 positions from the first 50,000 positions in **a** and display them as pink balls in Li$_6$PS$_5$Cl structure.

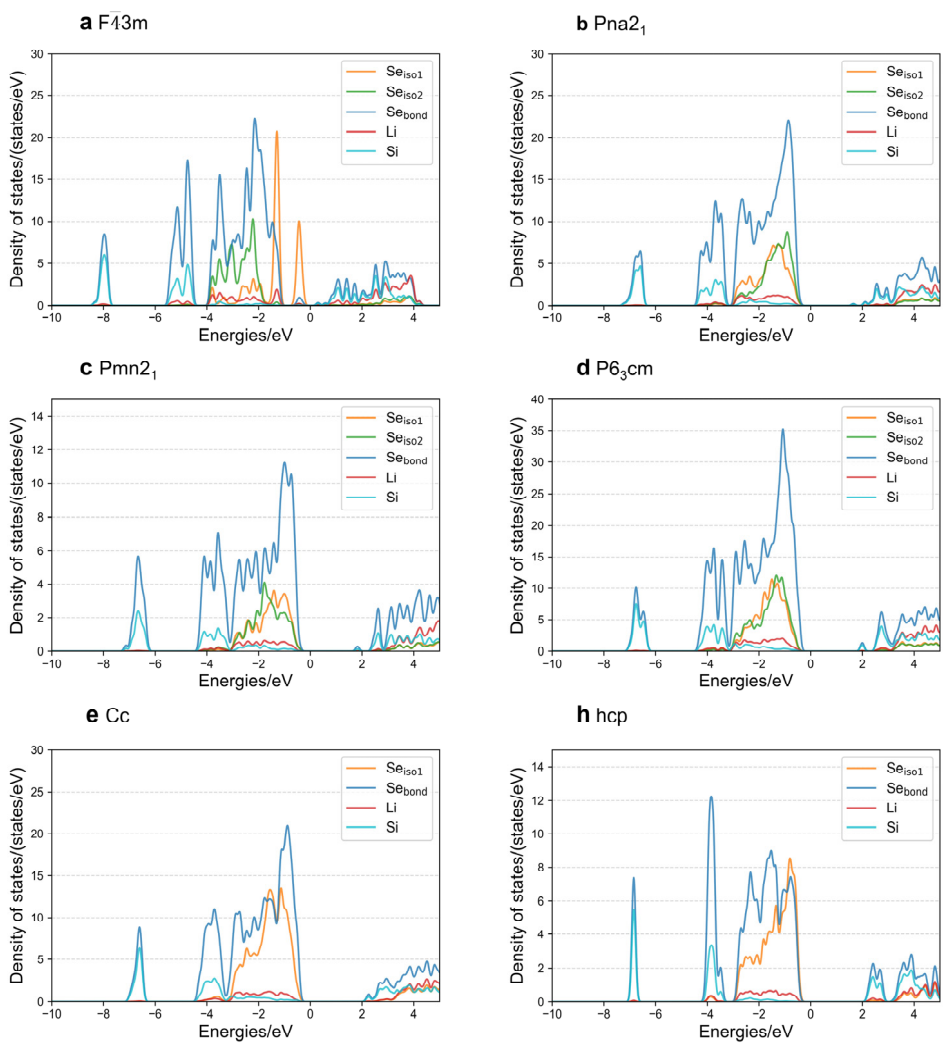

**Supplementary Figure S2. Partial density of electronic states (PDOS) of Li$_8$SiSe$_6$ structures.**

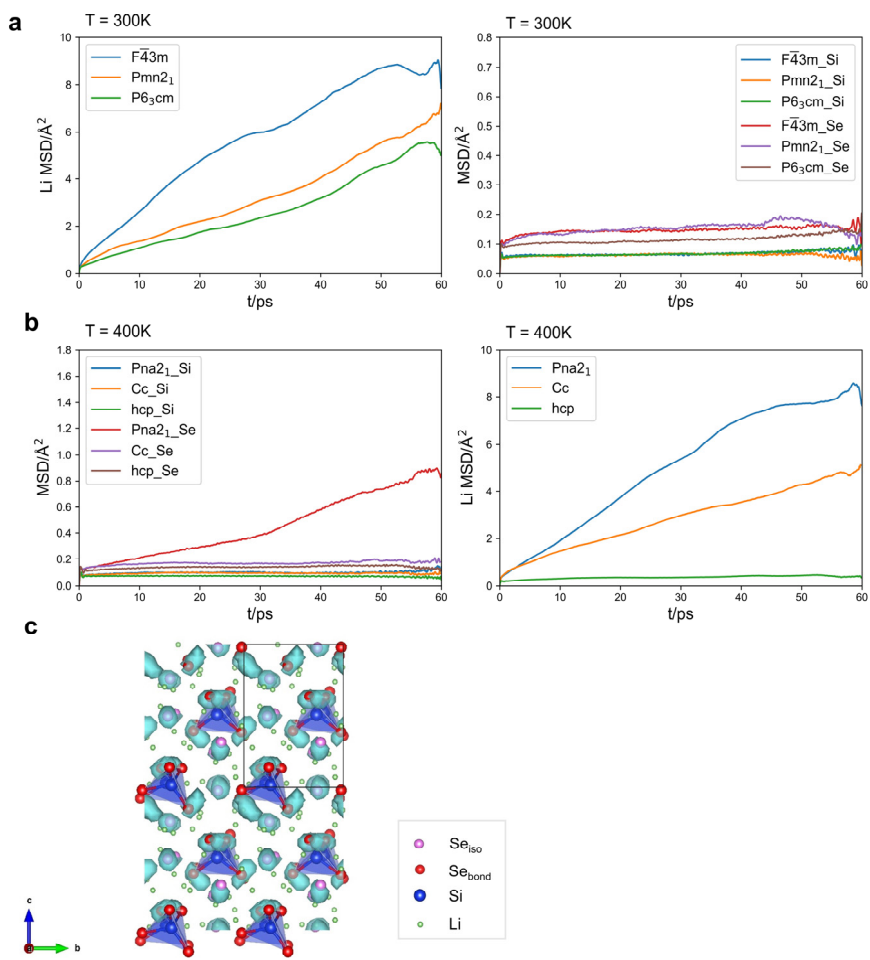

**Supplementary Figure S3. MSDs of Li$_8$SiSe$_6$'s structures calculated by AIMD.**
**a** MSDs of Li$_8$SiSe$_6$ in the space group of F$\bar{4}$3m, Pmn2$_1$ and P6$_3$cm at 300K. **b** MSDs of Li$_8$SiSe$_6$ in the space group of Pna2$_1$, Cc and hcp at 400K. **c** Se atoms' pathway in Li$_8$SiSe$_6$_Pna21 at 400K. In **a** and **b**, it is showed that Si and Se atoms are stable with no melting phenomenon because there is no movement of them was observed except Se in Li$_8$SiSe$_6$_Pna2$_1$ at 400K. For Li$_8$SiSe$_6$_Pna2$_1$ at 400K, it can be seen in **c** that the Se$_{bond}$'s rotation is responsible for the high MSD and Se$_{iso}$ has no movement during AIMD simulation, which indicates that the structure is stable.

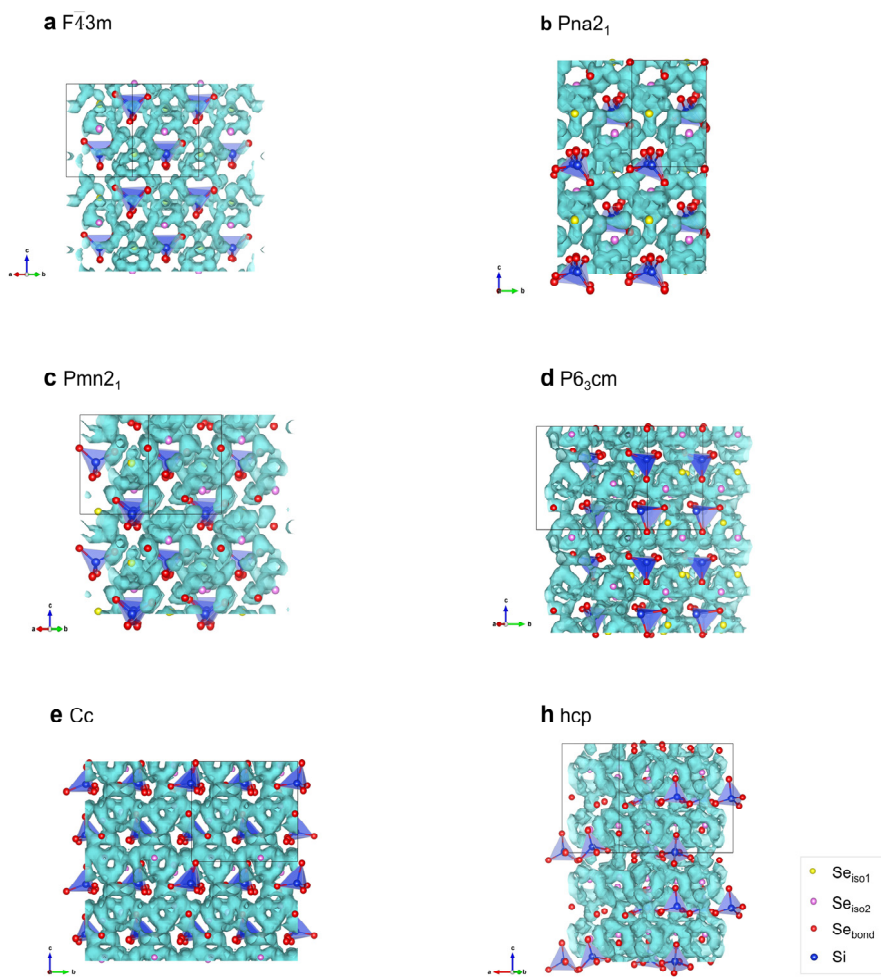

**Supplementary Figure S4. Li diffusion pathway of $Li_8SiSe_6$'s structures calculated by AIMD.** The Li atoms are hidden.

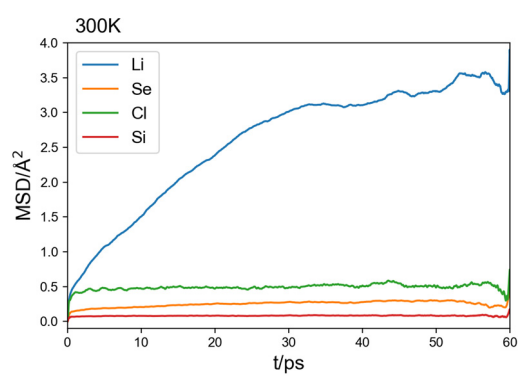

**Supplementary Figure S5. MSDs of Li$_7$SiSe$_5$Cl at 300K calculated by AIMD.**

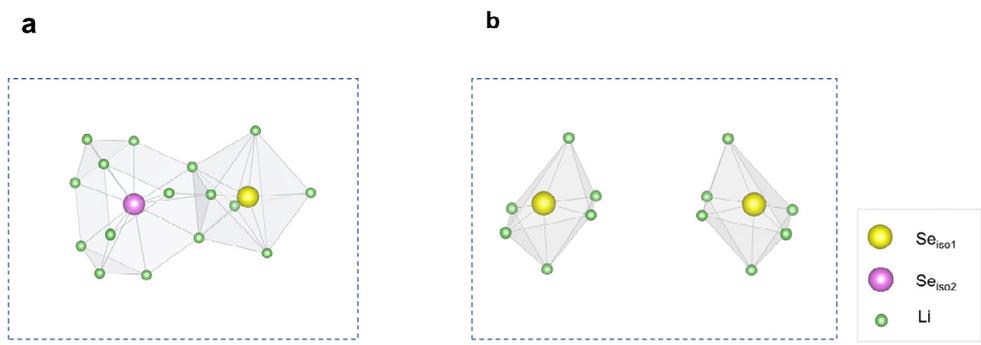

**Supplementary Figure S6.** Li sites' arrangement around adjacent $Se_{iso}$ anions in $Li_8SiSe_6\_F\bar{4}3m$ (a) and $Li_7SiSe_5Cl$ (b).

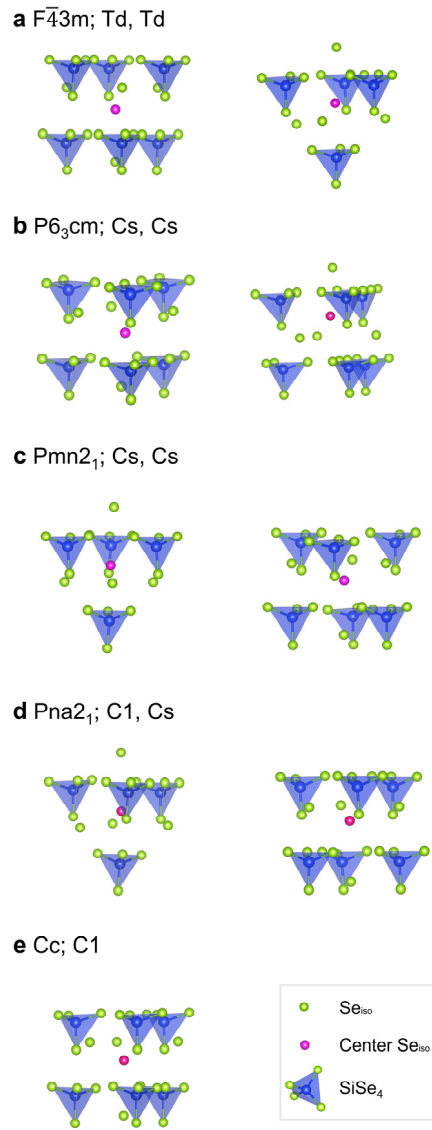

**Supplementary Figure S7. Isolated anion's local structure and its point group in $Li_8SiSe_6$'s structures.**

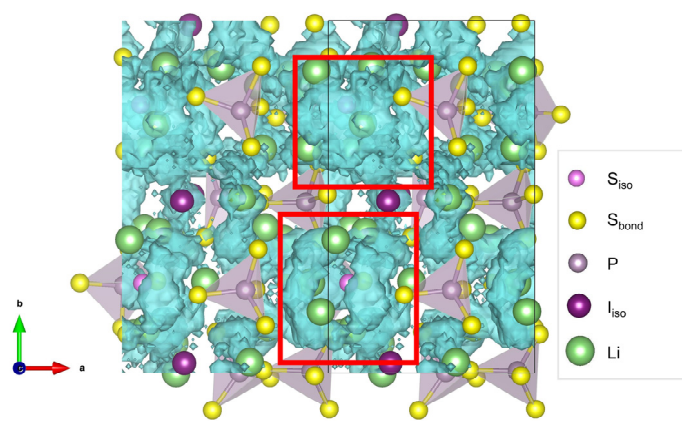

**Supplementary Figure S8. Li diffusion pathway of Li$_7$P$_2$S$_8$I calculated by AIMD.** The red boxes reveal Li cages around S$_{iso}$ and there is no migration of Li observed around I$_{iso}$ which means no Li cage formed around I$_{iso}$.

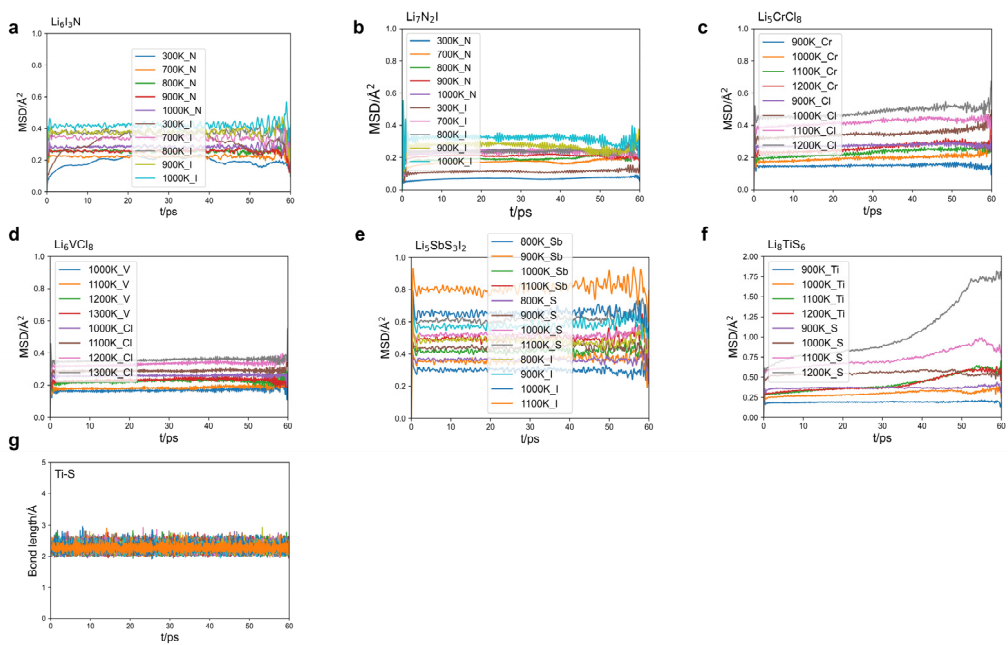

**Supplementary Figure S9. MSDs of atoms except Li in new conductors calculated by AIMD.** In **a-b**, it is showed that the non-Li atoms are stable with no melting phenomenon because there is no movement of them was observed except S in $Li_8TiS_6$ at 1200K. For $Li_8TiS_6$ at 1200K, it can be seen in **g** that the bond lengths between Ti and S remain unchanged, so the S's rotation is responsible for the high MSD and S has no movement during AIMD simulation, which indicates that the structure is stable.

**Supplementary Table S1. Other new conductors selected by high-throughput search.**

| Structure | Origination | Isolated anions' type |
|---|---|---|
| $Li_6NBr_3$ | icsd-083379 | N |
| $Li_4ClN$ | mp-29149 | N |
| $Li_5NBr_2$ | mp-29025 | N |
| $Li_{10}N_3Br$ | mp-28989 | N |
| $Li_8N_2Se$ | mp-973793 | N |
| $Li_9NS_3$ | mp-557964 | N |
| $Li_6NiCl_8$ | mp-1211124 | Cl |

**Supplementary Table S2. Total number of atoms in primitive cells and supercells.**

| Structure | Total Number of Atoms in Primitive Cell | Total Number of Atoms in Supercell |
|---|---|---|
| $Li_8SiSe_6\_F\bar{4}3m$ | 60 | 60 |
| $Li_8SiSe_6\_Pna2_1$ | 60 | 60 |
| $Li_8SiSe_6\_Pmn2_1$ | 30 | 60 |
| $Li_8SiSe_6\_P6_3cm$ | 90 | 90 |
| $Li_8SiSe_6\_Cc$ | 60 | 180 |
| $Li_8SiSe_6\_hcp$ | 30 | 120 |
| $Li_7SiSe_5Cl$ | 56 | 56 |
| $Li_7P_2S_8I$ | 36 | 72 |
| $Li_6NI_3$ | 40 | 40 |
| $Li_7N_2I$ | 80 | 80 |
| $Li_5CrCl_8$ | 14 | 70 |
| $Li_6VCl_8$ | 15 | 75 |
| $Li_5SbS_3I_2$ | 44 | 88 |
| $Li_8TiS_6$ | 30 | 120 |

**Supplementary Note 1. The calculation process of Euclidean distance of the Atom-Centered Symmetry Functions (ACSFs) descriptors between Li lattice sites and positions covering the entire structure in $Li_6PS_5Cl$.**

Because the framework structure composed of P, S, and Cl atoms forms the local chemical environments of Li, we only consider the impact of P, S, and Cl atoms on the chemical environment of Li. We calculated the descriptors of all Li lattice sites in the $Li_6PS_5Cl$ structure and found that the distances among them are all on the $10^{-14}$ order, indicating that all Li in the structure have the same chemical environment and are equivalent. In order to calculate the descriptors for each possible Li position of the framework structure, we first selected 1,000,000 positions equidistant. Then we discarded the lattice points with a distance of less than 2.2 Å from the framework atoms, leaving 275,315 positions. Finally, we calculated the descriptors for each position and the distance of descriptors from Li lattice sites to these positions. The smaller distance of ACSFs descriptor means the closer local environment between the position and the Li lattice sites. For example, the figure 5,000 in Supplementary Figure S1a means that 5000 positions on the left side of the red line have the most similar local environment as Li lattice sites. And we also selected 1,500 locations with very similar chemical environments to Li sites and found that they are almost all distributed around the isolated anions centered Li cage, which proves that the chemical environment and energy of the cage around isolated anions are very similar, and this also means the generation of frustration.
Therefore, we have demonstrated through calculations of different sites' descriptors that isolated anions can promote the frustration and calculating descriptors can be a universal method to determine whether a structure is experiencing frustration.

**Supplementary Note 2. The calculation process of vacancy formation energy.**
Due to the chaotic distribution of Li in structures, even Se at the same Wyckoff position will have different numbers of bonded Li atoms. So the vacancy formation energy of Se is determined by calculating the average value of all atoms of the corresponding type ($Se_{bond}$, $Se_{iso1}$ and $Se_{iso2}$).